\documentstyle[twoside,fleqn,espcrc2]{article}

\newcommand{\AmS}{{\protect\the\textfont2
  A\kern-.1667em\lower.5ex\hbox{M}\kern-.125emS}}

\input epsf.sty
\newdimen\psfigsize
\def\psfigure#1 #2 #3 #4 #5{
    \begin{figure}[tbh]
    \vbox{
    \null\vskip-0.2in\hskip#2\epsfxsize=#1 \epsfbox[0 0 4096 4096]{#4}
    \vskip -0.5in
    \caption {#5 \label{#3}}
     \vskip -0.15in}
    \end{figure}
}
\def\psfigtop#1 #2 #3 #4 #5{
    \begin{figure}[t]
    \vbox{
    \null\vskip-0.2in\hskip#2\epsfxsize=#1 \epsfbox[0 0 4096 4096]{#4}
    \vskip -0.3in
    \caption {#5 \label{#3}}
     \vskip -0.15in}
    \end{figure}
}

\def\psoddfigure#1 #2 #3 #4 #5 #6{
    \begin{figure}[tbh]
    \vbox{
    \null\vskip-0.2in\hskip#3\epsfxsize=#1 \epsfbox[0 0 4096 4096]{#5}
    \vskip -#1 \vskip #2 \vskip 10truept
    \vskip -0.2in
    \caption {#6 \label{#4}}
    \vskip 0.0truein plus0.2truein}
    \end{figure}
}

\title{Update on the hadron spectrum with two flavors of staggered quarks}

\author{ Claude~Bernard,\address{{\vskip-0.10in{\hskip 0.07in Department
of Physics, Washington University, St.~Louis, MO 63130, USA}}} 
Tom Blum, 
\address{Department of Physics, Brookhaven National Lab, Upton, NY 11973, USA}
Thomas~A.~DeGrand,
\address{Physics Department, University of Colorado,
Boulder, CO 80309, USA} 
Carleton~DeTar,
\address{Physics Department, University of Utah, Salt Lake City, UT
84112, USA} 
Steven~Gottlieb,
\address{Department of Physics, Indiana University, Bloomington, IN
47405, USA} 
Urs M.~Heller,
\address{SCRI, Florida State University, Tallahassee, FL 32306-4130, USA}
Jim~Hetrick, $\,\null^{\rm a}$
\thanks{Current address, Department of Physics, University of the
Pacific, Stockton, CA 95211, USA}
Craig McNeile, $\,\null^{\rm c}$
Kari~Rummukainen,
\address{Universit\"at Bielefeld, Fakult\"at f\"ur Physik, Postfach
100131, D-33501 Bielefeld, Germany} 
Robert~Sugar
\address{Department of Physics, University of California, Santa Barbara,
CA 93106, USA} 
\thanks{presented by Robert Sugar}
and Doug~Toussaint
\address{Department of Physics, University of Arizona, Tucson, AZ 85721, USA} 
}
       
\begin{document}

\begin{abstract}
We present an update on the MILC Collaboration's light hadron spectrum
calculation with two flavors of dynamical, staggered quarks. Results are 
presented for gauge couplings 5.30, 5.415, 5.50 and 5.60, with a range of 
quark masses for each value of the coupling. We present extrapolations of
$m_N/m_\rho$ to the continuum limit for fixed values of $m_\pi/m_\rho$
including the physical one.
\end{abstract}

\maketitle


Over the past few years the MILC Collaboration has been engaged in
a series of spectrum calculations with dynamical quarks and in
the quenched approximation, using both standard and improved
actions\cite{EARLIER}. One objective of this work is to obtain sufficient
control over systematic errors so that one can make reliable
extrapolations to the limits of zero lattice spacing and physical
quark masses. Another objective is to make detailed comparisons
between the spectra with quenched and dynamical quarks, and
with standard and improved actions. Clearly, to reach these
objectives requires high statistics calculations for a sufficiently
wide range of lattice spacings and quark masses to enable us to make
extrapolations to the continuum and chiral limits. It also requires
that these calculations be carried out on large enough lattices
to avoid finite size effects.

In this note we provide an update on our results for two flavors
of dynamical, staggered quarks using the standard gauge and quark
actions. Results for the quenched approximation and with improved
actions are presented elsewhere in these proceedings\cite{STEVE}.
We have carried out simulations at four values of the gauge coupling,
$6/g^2=5.30$, 5.415, 5.50 and 5.60 using at least four values of
the quark mass at each coupling. Lattices were generated with the
refreshed hybrid molecular dynamics algorithm. Once the lattices
were equilibrated the light hadron spectrum was measured every fifth
molecular dynamics time unit. Lattices were saved after every
tenth time unit for use in other projects. With the exeption of
a few runs at large quark mass, measurements were made on at least 
four hundred lattices at each values of the gauge couping and quark
mass.

As one would expect, the hadron masses evaluated in lattice units are
strongly dependent on the gauge coupling and bare lattice quark
mass, $am_q$. Here $a$ is the lattice spacing. The one exception is the 
mass of the Goldstone pion 
which is weakly dependent on the gauge coupling. In Fig.~1,
we plot the $\rho$ mass as a function of the bare quark mass
for the two weakest values of the gauge coupling that we have
studied. For comparison we also include in this figure results 
from quenched calculations with staggered quarks at gauge couplings 
5.70, 5.85 and 6.15.  Notice that for fixed values of the gauge 
coupling, $am_\rho$ decreases more rapidly for decreasing $am_q$ in 
full QCD than in the quenched approximation. This is due to the
dependence of the lattice spacing on the quark mass in full QCD.

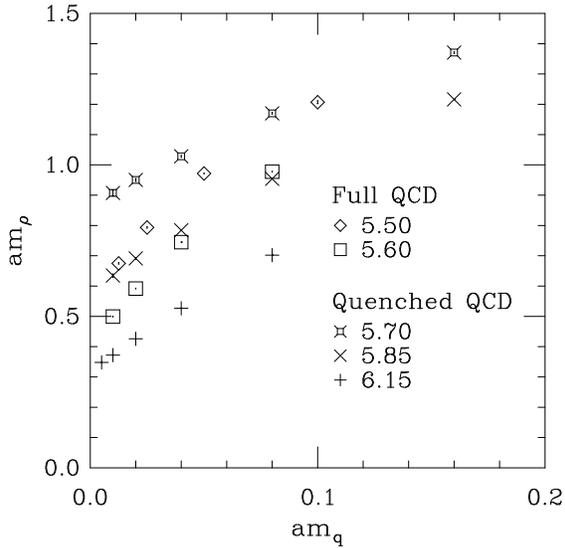
\psfigure 3.0in -0.15in {FIG1} {fig1.ps} {The $\rho$ mass as a function
of the quark mass for full and quenched QCD.} 

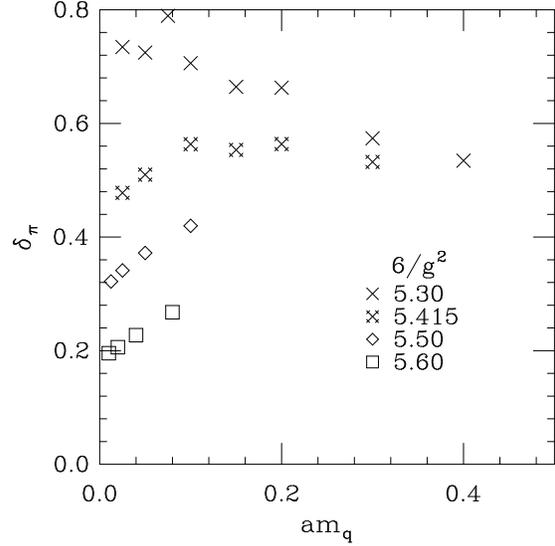
\psfigure 3.0in -0.15in {FIG2} {fig2.ps} {$\delta_\pi$ as a function
of $am_q$ for full QCD.}

An important question regarding any simulation with staggered
quarks is the extent to which flavor symmetry is restored
as the lattice spacing and quark mass are decreased. One
measure of flavor symmetry violation is the quantity
$\delta_\pi =(m_{\pi_2}^2 - m_{\pi}^2) / m_{\rho}^2$.
In Fig.~2, we plot $\delta_\pi$ as a function of $am_q$
for each of the couplings we have studied. The trends are
as expected. The value of $\delta_\pi$ for $6/g^2=5.60$
is close to, but slightly below that found in the quenched
approximation for gauge coupling 5.85.

$$
    \vbox{
    \null\vskip-0.2in\hskip-0.30in \epsfxsize=3.0in \epsfbox[0 0 4096 4096]{fig3.ps}
    \vskip -0.3in
    \vskip 0.0truein plus0.2truein}
$$
\vskip 0.1in
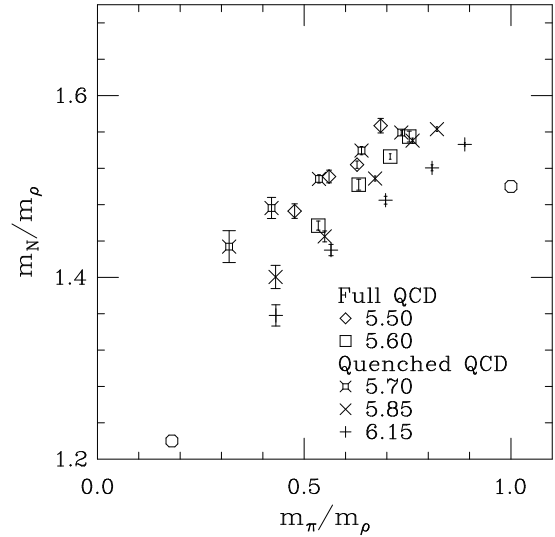
{\noindent Figure 3. The Edinburgh plot for full and quenched QCD.}
\vskip 0.2in

In Fig.~3, we show the Edinburgh plot for our two weakest
gauge couplings using the Goldstone pion in the ratio
$m_\pi/m_\rho$. Once again we include quenched results for gauge
couplings 5.70, 5.85 and 6.15 in this graph. Here too, the 
full QCD results at 5.60 are quite close to
the quenched ones at 5.85.

Because we have carried out calculations with a number of
quark masses at each value of the coupling we have studied,
we can perform fits to the hadron masses as a function of
$am_q$ in order to extrapolate to the chiral limit and to
interpolate for comparisons with quenched and improved action
calculations. We are analyzing a variety of fitting functions.
Here we present results from fits of the nucleon and rho masses
to the form
$$
m=m_0+\alpha m_q+\beta m_q^{3/2}+\gamma m_q^2.
$$
In Fig.~4 we show $m_N/m_\rho$ for the four values of the gauge
coupling we have studied, in each case interpolating to the
value of $am_q$ for which $m_\pi/m_\rho=0.5$. The $x$-axis in
this figure is $(am_\rho)^2$ interpolated to the same value
of the quark mass. This quantity gives a measure of the square
of the lattice spacing.

\addtocounter{figure}{1}

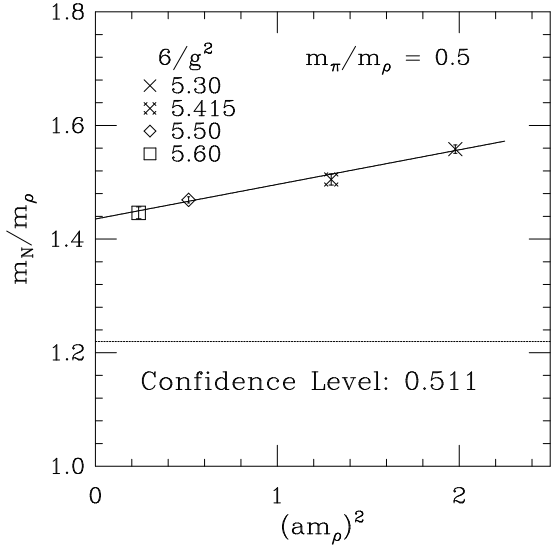
\psfigtop 3.0in -0.15in {FIG4} {fig4.ps} {$m_N/m_\rho$ as a
function of the square of the lattice spacing for $m_\pi/m_\rho=0.5$}

Fig.~5 is a repeat of Fig.~4, this time with extrapolations
to the value of the quark mass for which 
$m_\pi/m_\rho$ takes on its experimental
value, 0.1753. The errors for the individual points in Figs.~4 and 5 
were determined by a jackknife analysis.  The solid lines in
these figures are fits to the four interpolated or extrapolated points
of the form $\alpha + \beta (am_\rho)^2$,
where $\alpha$ and $\beta$ are fit parameters.
The confidence levels of these fits
are shown in the figures. The fits allow an extrapolation
of $m_N/m_\rho$ to the continuum limit for fixed values of
$m_\pi/m_\rho$, and they provide support of the expectation that 
for staggered quarks the leading corrections are of order $a^2$.
These results are quite encouraging; however, a better understanding
of the chiral extrapolation, as well as additional calculations at
weaker coupling and smaller quark masses are needed in order to 
obtain definitive results.

This work was supported by the Department of Energy and the
National Science Foundation.
Computations were carried out at ORNL, PSC,
NCSA, IU, SDSC, CTC and MHPCC.

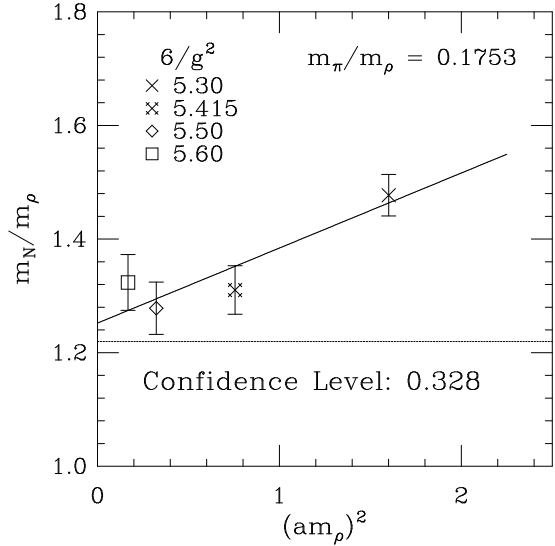
\psfigtop 3.0in -0.15in {FIG5} {fig5.ps} {$m_N/m_\rho$ as a
function of the square of the lattice spacing with $m_\pi/m_\rho$ fixed at its
experimental value.}

\end{document}